\documentclass[11pt,a4paper]{article}
\pdfoutput=1
\usepackage{jheppub} % for details on the use of the package, please
\usepackage{graphicx}
\usepackage{amsmath}
\usepackage{indentfirst}
\usepackage{epsf}
\usepackage{epsfig}
\usepackage[makeroom]{cancel}
\usepackage{caption}
\usepackage{amssymb}
\newcommand{\be}{\begin{equation}}
\newcommand{\ee}{\end{equation}}
\newcommand{\bear}{\begin{eqnarray}}
\newcommand{\ear}{\end{eqnarray}}
\newcommand{\ba}{\begin{array}}
\newcommand{\ea}{\end{array}}

\makeatletter

\def\compoundrel#1\over#2{\mathpalette\compoundreL{{#1}\over{#2}}}
\def\compoundreL#1#2{\compoundREL#1#2}
\def\compoundREL#1#2\over#3{\mathrel
         {\vcenter{\hbox{$\m@th\buildrel{#1#2}\over{#1#3}$}}}}
\makeatother

\preprint{
  \begin{flushright}
    INR-TH-2017-019\\
    DESY-17-132
  \end{flushright}
  }
\title{Low scale supersymmetry at the LHC with jet and missing energy signature}

\author[a,b]{S.V. Demidov,}
\author[a,c,d]{I.V. Sobolev}

\affiliation[a]{Institute for Nuclear Research of the Russian Academy of Sciences,\\ 60th October Anniversary prospect 7a, Moscow 117312, Russia}
\affiliation[b]{Moscow Institute of Physics and Technology, \\ Institutsky per. 9, Dolgoprudny 141700, Russia}
\affiliation[c]{M.V. Lomonosov Moscow State University, \\ Vorobjevy Gory, 119991, Moscow, Russia}
\affiliation[d]{DESY, Notkestra{\ss}e 85, 22607 Hamburg, Germany}

\emailAdd{demidov@ms2.inr.ac.ru}
\emailAdd{sobolev.ivan@physics.msu.ru}

\abstract{If supersymmetry is broken at TeV scale, particles from sector responsible for supersymmetry breaking - goldstino and sgoldstinos - can reveal themselves already at the LHC experiments. We discuss bounds on supersymmetry breaking
scale from the LHC searches for events with a jet plus missing momentum
signature focusing on the case of TeV scale sgoldstinos. We show
that contribution of light sgoldstinos to the cross section of
of gravitino pair production with a jet can be sizable and the
bounds on the gravitino mass can be stronger by up to a factor of 2 as compared to those obtained in the heavy sgoldstino limit. We compare these bounds
on parameters of the model to those obtained with the results of ATLAS and
CMS searches for dijet resonances.
}

\keywords{Supersymmetry, Phenomenology, Beyond the Standard Model, Hadronic Colliders, Supergravity Models, Supersymmetry Breaking}

\begin{document}
\maketitle
\toccontinuoustrue

 \section{Introduction}
Supersymmetry is among the most interesting and well motivated
Standard Model (SM) extensions. It offers solutions to different SM
problems including dark matter, naturalness and gauge couplings unification
~\cite{Martin:1997ns}. If supersymmetry is indeed a true symmetry of Nature it should be
broken in some way. In the standard approach the MSSM lagrangian is
supplied with so-called soft terms which break supersymmetry
explicitly. All mechanisms of spontaneous supersymmetry breaking which explain existence of these terms inevitably use so-called ``hidden sector''~\cite{Martin:1997ns} where spontaneous supersymmetry breaking occurs. The information that
supersymmetry is broken is then transmitted to the observable sector
by means of some messenger fields. In the simplest case ``hidden sector'' can be decribed by one singlet chiral superfield $\left(\phi, \psi,
F_{\phi}\right)$ \footnote{See~\cite{Brignole:1996fn,Luty:1996um} for
  extensive discussion of possibility of charged ``hidden sectors''.}.
Here $\psi$ is a massless Goldstone fermion called goldstino, $\phi$
is its scalar superpartner, sgoldstino, and $F_\phi$ is an auxiliary
field. The scale of supersymmetry breaking is often denoted as
$\sqrt{F}$, where $F$ -- is a vacuum expectation value of the
auxiliary field $F_{\phi}$. In supergravity theories, where supersymmetry is promoted to a local symmetry, goldstino becomes a longitudinal part of spin-$\frac{3}{2}$ particle -- gravitino $\tilde{G}$ -- due to Super-Higgs mechanism~\cite{Volkov:1973jd,Deser:1977uq,Cremmer:1978iv}. Gravitino mass $m_{3/2}$ is then completely determined by the scale of supersymmetry
breaking $\sqrt{F}$ 
\[
m_{3/2} = \sqrt{\frac{8\pi}{3}} \frac{F}{M_{\rm Pl}}.
\]

If gravitino is sufficiently heavy, e.g. its mass is of order of
electroweak scale $v$, then $\sqrt{F} \sim \sqrt{M_{\scriptstyle \rm{Pl}}~v} \sim
10^{10} - 10^{11}~\rm{GeV}$ and the effective theory near the
electroweak scale is just the MSSM endowed with the soft terms. However models  with relatively light gravitino (i.e. with a mass of several keV) also exist and are phenomenologically viable. In particular, this is the case of models with gauge mediated SUSY breaking~\cite{Dine:1981za,Dimopoulos:1981au,Dine:1981gu,Dine:1982zb,Nappi:1982hm,AlvarezGaume:1981wy,Dine:1993yw,Dine:1994vc,Dine:1995ag} and no-scale supergravity~\cite{Ellis:1984kd,Ellis:1984xe,Lahanas:1986uc} models where $\sqrt{F}\sim 10 - 100~\rm{TeV}$. However in this study we will focus on the case when $\sqrt{F}$ lies even lower, i.e. when it is of order of several TeVs, which means that gravitino mass lies in sub-eV range. This can be realized in warped~\cite{Gherghetta:2000qt,Gherghetta:2000kr} SUSY models and composite~\cite{Luty:1996um} models with charged ``hidden sector''. Although there is not a lot of models of the latter class their study is interesting for the following reasons. First of all, the couplings of particles from the ``hidden sector'' with other SM fields scale as powers of $1/F$. In this way, the smaller $F$ the stronger manifestation of new physics effect would be. Second, perturbative unitarity constrains possible masses of sgoldstinos~\cite{Brignole:1996fn}. In particular, their masses cannot be significantly larger than $\sqrt{F}$ scale. If $\sqrt{F}$ is of order of several TeVs than sgoldstino mass can lie in a range, the LHC~\cite{Gorbunov:2002er,Demidov:2004qt,Asano:2017fcp} and forthcoming experiments~\cite{Astapov:2015otc,Astapov:2016koq} will be sensitive to. In the regime when typical collision energy is much large than the gravitino mass, i.e. $E \gg m_{3/2}$, equivalence theorem~\cite{Fayet:1977vd,Casalbuoni:1988kv,Casalbuoni:1988qd} allows to approximate gravitino interactions by interactions of its longtitudial part
\[
\tilde{G} \sim \frac{\partial_{\mu} \psi}{m_{3/2}}.
\]
Hence the effective lagrangian describing interactions of ultralight gravitino with the MSSM fields can be constructed using properties of goldstino. In this study we deal with sub-eV gravitinos being produced at the LHC at 8 TeV. Hence throughout this paper we do not make any difference between goldstino and gravitino.

Search for a signal from physics beyond the Standard Model is one of the top priority goals of the LHC experiments at the moment. Many different signatures are
thoroughly scrutinized to constrain models of new physics. Although the LHC experiments do not have any significant evidence of a
signal from SUSY, this class of models remains the most extensively studied. Different aspects of collider phenomenology of low scale
supersymmetry breaking have been studied  for instance in~\cite{Demidov:2004qt,Perazzi:2000ty,Perazzi:2000id,Dudas:2012fa,Gorbunov:2000ht,Brignole:2003cm,Petersson:2012nv}. Gravitino production is one of the most distinct signatures of this
setup~\cite{Dicus:1991wj,Shirai:2009kn,Mawatari:2014cja,Ellis:1996aa,Klasen:2006kb,Brignole:1998me,Brignole:1997sk,Petersson:2012dp}.
Due to its $R$-odd nature, LSP goldstino would be always produced in
pairs. In this paper we discuss jet-plus-missing-momentum signature of
the process of gravitino pair production at the LHC. It has been recently
carefully studied in~\cite{Maltoni:2015twa} where authors redid jet plus
transverse missing energy analysis performed by ATLAS with a part of run-I data
to constrain gravitino mass or equivalently supersymmetry breaking
scale. However, in that work the limit of very heavy sgoldstino  
(with mass about 20~TeV) was considered. The goal of our work is to
include possibility of TeV scale sgoldstino contribution.

The paper is organized as follows. In Section~\ref{sec1} we describe a minimal model with goldstino
supermultiplet and introduce possible interactions of this
supermultiplet to the SM fields. In Section~\ref{sec2} we
discuss processes contributing to the gravitino pair production in
proton-proton collisions and in particular discuss contribution of light
sgoldstinos. Then we describe bounds on the parameters of the
model from missing energy signature analysis performed in our work
using run-I data of the LHC experiments and
describe comparison of the obtained results to those obtained using dijet
searches. Section~\ref{conclusions} contains our conclusions. 

\section{Model description}
\label{sec1}
\noindent
We choose a simple Polonyi model~\cite{Polonyi:1977pj} to describe dynamics of the ``hidden
sector'' 
\begin{equation}
\label{mdl:lgr1}
\mathcal{L}_{\Phi} = \int \, d^2 \theta \, d^2 \bar{\theta} \left(
\Phi^{\dagger}\Phi + \tilde{K} (\Phi^{\dagger},\Phi)\right) + \left(
\int d^2 \theta F \Phi + \text{h.c.} \right),  
\end{equation}
where $\Phi = \phi+\sqrt{2} \theta \psi + \theta^2 F_{\phi}$ is goldstino
chiral superfield and non-canonical part of the Kahler potential
$\tilde{K}$ can be chosen in the form\footnote{Here $~K(\phi,\phi^*) = \phi^*
  \phi + \tilde{K}(\phi,\phi^*)$. Also we use shorthand notations 
  \[
  W_{\phi} = \frac{\partial W(\phi)}{\partial \phi},~\overline{W}_{\phi^*}
  = \frac{\partial \overline{W}(\phi^*)}{\partial \phi^*},~K_{\phi \phi^*}
  = \frac{\partial^2 K(\phi, \phi^*)}{\partial \phi \partial \phi^*}, 
  \]
}
\begin{equation}
\label{mdl:lgr2}
\tilde{K}(\Phi^{\dagger},\Phi) = - \frac{m_{s}^2+m_{p}^2}{8F^2}
\left(\Phi^{\dagger}\Phi \right)^2 - \frac{m_{s}^2-m_{p}^2}{12F^2}
\Phi^{\dagger} \Phi \left(\Phi^2+(\Phi^{\dagger})^2\right).
\end{equation}
Non-renormalizable operators in this expression can be  generated by
integrating out some heavy states in the microscopic theory.
The scalar potential of the model has the following form\footnote{Hereafter we assume
  that all parameters in the lagrangian are real. In particular, $F
  = F^*$.}  
\begin{equation}
  \label{mdl:lgr3}
  V(\phi^{*},\phi) = W_{\phi} \overline{W}_{\phi^*} K_{\phi \phi^*}^{-1} =
  \frac{F^2}{\displaystyle 1-\frac{m_s^2+m_p^2}{2F^2} \vert \phi \vert^2 -
    \frac{m_s^2-m_p^2}{4F^2}(\phi^2+(\phi^*)^2)} =
  \frac{F^2}{\displaystyle 1-\frac{m_s^2 s^2+m_p^2 p^2}{2F^2}} .
\end{equation}
This potential has a local minimum where $\langle s \rangle = \langle
p \rangle = 0$ and its expansion around local minimum gives proper
mass term for CP-even (s) and CP-odd (p) sgoldstinos 
\[
V(s,p) = F^2+\frac{m_s^2 s^2}{2}+\frac{m_p^2 p^2}{2} +
\mathcal{O}\left(\frac{1}{F^2}\right) 
\]
In particular, if $m_{s} \neq m_{p}$ scalar and pseudoscalar components of
$\phi = \frac{1}{\sqrt{2}}\left(s+ i p\right)$ get different masses.
Furthermore, we see that vacuum has positive energy density, $\langle V
\rangle = F^2 > 0$, so supersymmetry is spontaneously
broken. Auxiliary fields $F_{\phi}$ and $F_{\phi}^*$ acquire non-zero
vacuum expectation values 
\begin{equation}
  \label{mdl:lgr4}
  \left. F_{\phi} \right|_{\rm{vac}}= \left. \left(\frac{1}{2}
  K_{\phi^* \phi \phi} (\psi \psi) - \overline{W}_{\phi^*} \right) K_{\phi
    \phi^*}^{-1} \right|_{\rm{vac}} = -F 
\end{equation}
\noindent
As was already mentioned in the Introduction the quantity
$\sqrt{F}$ has dimension of energy and has the meaning of energy scale
of supersymmetry breaking. 

Interaction of goldstino superfield $\Phi$ with other particles
of MSSM, i.e.``visible'' sector, can be introduced by making use
of spurion technique. Recall that soft terms in lagrangian of the MSSM
formally can be written in ``manifestly supersymmetric'' way by using the
following operators  
\[
\left. S^{\dagger} S \Phi^{\dagger} \Phi \right|_{D}, \left. \mu S
\Phi^2 \right|_{F}, \left. S \Phi^3 \right|_{F}, \left. S W^{\alpha}
W_{\alpha} \right|_{F}, 
\]
where $S = \theta^2 m_{\rm soft}$ is a spurion superfield, $\Phi$ is a matter
chiral superfield and $W^{\alpha}$ is a gauge field strength. We
promote spurion $S$ to the dynamical field $\Phi$, containing goldstino
and sgoldstinos by the following rule 
\[
S \to m_{\rm soft} \frac{\Phi}{F},
\]
In what follows we will be interested in interactions between invisible and QCD sectors of the MSSM. Spurion method allows to describe this interaction by the following lagrangian
\begin{eqnarray}
  \label{mdl:lgr5}
  \begin{aligned}
    & \mathcal{L}_{\Phi-\rm{vis}} =  - \int d^2 \theta d^2
    \theta^{\dagger} \frac{M_{\tilde{q}_{L},ij}^2}{F^2} \Phi^{\dagger}
    \Phi Q_{L,i}^{\dagger} e^{2gV} Q_{L,j} - \int d^2 \theta d^2
    \theta^{\dagger} \frac{M_{\tilde{u}_{R},ij}^2}{F^2} \Phi^{\dagger}
    \Phi U_{R,i}^{\dagger}  e^{2gV} U_{R,j} -\\ 
    & - \int d^2 \theta d^2 \theta^{\dagger}
    \frac{M_{\tilde{d}_{R},ij}^2}{F^2} \Phi^{\dagger} \Phi
    D_{R,i}^{\dagger}  e^{2gV} D_{R,j} + \left( \frac{M_{3}}{2F} \int d^2
    \theta~\Phi W^{\alpha} W_{\alpha} + \rm{h.c.} \right). 
    \end{aligned}    
\end{eqnarray}  
Here $Q_{L,i}$, $U_{R,i}$ and $D_{R,i}$ are (s)quarks superfields, $W^{\alpha}$ is a strength tensor superfield containing gluons and gluinos, $M_{\tilde{q}_{L},ij}$, $M_{\tilde{u}_{R},ij}$, $M_{\tilde{d}_{R},ij}$ and $M_{3}$ are respective soft masses and indices $i,j$ run over generations of quarks. Apart from soft quark and gluino masses the above lagrangian generates a tower of interactions of goldstino with the MSSM fields. In particular it contains the following operators obtained by expansion
of component field lagrangian in powers of $1/F$
\begin{eqnarray}
  \label{mdl:lgr6}
  \begin{aligned}
    & \mathcal{L}_{\psi-\rm{vis}} \supset  \frac{M_{3}}{4\sqrt{2}F}
    \bar{\psi} \left [ \gamma^{\mu}, \gamma^{\nu} \right ] \lambda^{a}
    F_{\mu \nu}^{a} - \frac{i M_{\tilde{d}_{R},ij}^2}{F}
    \left(\tilde{d}_{R,i}^{\dagger} \bar{\psi} P_{R} d_{j} -
    \bar{d}_{j} P_{L} \psi \tilde{d}_{R,i} \right) - \\ 
    & - \frac{i M_{\tilde{u}_{R},ij}^2}{F}
    \left(\tilde{u}_{R,i}^{\dagger} \bar{\psi} P_{R} u_{j} -
    \bar{u}_{j} P_{L} \psi \tilde{u}_{R,i} \right) + \frac{i
      M_{\tilde{q}_{L},ij}^2}{F} \left(\tilde{q}_{L,i}^{\dagger}
    \bar{\psi} P_{L} q_{j}  - \bar{q}_{j} P_{R} \psi \tilde{q}_{L,i}
    \right) - \\ 
    & - \frac{M_{\tilde{d}_{R},ij}^2}{F^2} (\bar{\psi} P_{R} d_{i})
    (\bar{d}_{j} P_{L} \psi) -  \frac{M_{\tilde{u}_{R},ij}^2}{F^2}
    (\bar{\psi} P_{R} u_{i}) (\bar{u}_{j} P_{L} \psi) -
    \frac{M_{\tilde{q}_{L},ij}^2}{F^2} (\bar{\psi} P_{L} q_{j})
    (\bar{q}_{i} P_{R} \psi) 
  \end{aligned}
\end{eqnarray}
which we will use in our study\footnote{We make use of four-component spinors in~\eqref{mdl:lgr5} -- \eqref{mdl:lgr7}. We've also redefined goldstino field
$\psi \to i \gamma_{5} \psi$ in~\eqref{mdl:lgr6}. This corresponds to redefinition of two-component spinor $\psi^{\alpha} = -i \psi^{\alpha}$.}. Here $\tilde{q}_{L,i}$, $\tilde{u}_{R,i}$ and $\tilde{d}_{R,i}$ denote  squarks and $\lambda^{a}$ denotes gluino field. It was shown in~\cite{Lee:1998aw} that in the case of negligible squark mixing lagrangian \eqref{mdl:lgr6} is equivalent to Goldberger-Treiman lagrangian up to some redefinition of fields
\begin{equation}
  \label{mdl:lgr5_a}
  \mathcal{L}_{GT} = \frac{1}{F} \partial_{\mu} \psi^{\alpha}
  J_{\alpha}^{\mu} + \rm{h.c.}, 
\end{equation}
with $J_{\alpha}^{\mu}$ being a supercurrent of the MSSM which contains all of the fermion-boson pairs from the visible sector. 
This equivalence means that both interaction lagrangian result in identical amplitudes for processes with a single 
external goldstino. Nonetheless, they give different answers for double goldstino production. Since in this study we are interested in
the latter case we utilize~\eqref{mdl:lgr6} together with relevant part of interaction lagrangian (see Eqs.~\eqref{mdl:lgr1}, \eqref{mdl:lgr2} and~\eqref{mdl:lgr5}) containing scalar and pseudoscalar
sgoldstinos 
\begin{equation}
  \label{mdl:lgr7}
  \mathcal{L}_{\phi-\psi} = \frac{m_s^2}{2\sqrt{2}F}s\bar{\psi}\psi +i
  \frac{m_p^2}{2\sqrt{2}F}p\bar{\psi}\gamma_{5} \psi - \frac{M_{3}}{2
    \sqrt{2}F} s F^{\mu \nu}_{a} F_{\mu \nu}^{a} +
  \frac{M_{3}}{2\sqrt{2}F} p F^{\mu \nu}_{a} \tilde{F}_{\mu \nu}^{a} 
\end{equation}
where $\tilde{F}_{\mu \nu} = \frac{1}{2} \epsilon_{\mu \nu \alpha \beta} F^{\alpha \beta}$ with $\epsilon_{0123} = +1$.

\section{Light gravitino production at the LHC: monojet signature}
\label{sec2}

In this Section we describe main subprocesses which result in jet-plus-missing-energy signature, discuss sgoldstino contribution 
and obtain bounds on parameter space of the model using the LHC run-I
data. To our knowledge the latest study in this direction was
performed in~\cite{Maltoni:2015twa} with ATLAS data at
$\sqrt{s}=8$~TeV and 10~fb$^{-1}$. In that study authors obtained
bounds on supersymmetry breaking scale $\sqrt{F}$ in the range
between $850$ and $1300$~GeV depending on the mass scale of superpartners and their
hierarchy. Our task is to extend this analysis and include possibility
of relatively light sgoldstinos. For the present study we make use
of CMS data at $\sqrt{s}=8$~TeV and 19.6~fb$^{-1}$ and perform leading
order (LO) analysis at parton level.

Three main signal subprocesses relevant for our study at LO are 1) gravitino
pair production in association with a parton (quark or gluon); 2)
production of single gravitino with squark or gluino; 3) SUSY QCD pair
production. Here it is assumed that squarks and gluino decay mainly as
$\tilde{q}(\tilde{g})\to q(g) + \psi$.
Corresponding decay widths look as 
\begin{equation}
  \label{decay1}
  \Gamma(\tilde{q}(\tilde{g}) \to q(g) + \psi) = \frac{M_{\tilde{q}(3)}^{5}}{16\pi F^2}.
\end{equation}
In what follows we assume that $M_{\tilde{q}}=M_{3}$ and these
decay modes are dominant. One can check that the latter assumption is 
actually valid for masses of squarks and gluinos and SUSY breaking
scale around TeV scale.
In general, if sgoldstinos have masses around TeV scale they can decay into a pair of MSSM particles and corresponding decay
widths are governed by relevant soft SUSY breaking parameters~\cite{Gorbunov:2001pd}.
However in the present analysis we consider only two decay channels: decay into a pair of gravitinos
\begin{equation}
  \label{decay2}  
  \Gamma(s(p) \to \psi \psi) = \frac{m_{s(p)}^{5}}{32 \pi F^2}  
\end{equation}
or a pair of gluons
\begin{equation}
  \label{decay3}  
  \Gamma(s(p) \to g g) = \frac{M_3^2 m_{s(p)}^{3}}{4 \pi F^2}.
\end{equation}
The reason is that these decay modes are indeed dominant for a typical hierarchy of soft
SUSY breaking parameters~\footnote{We disregard here possibility of sgoldstino mixing with the Higgs bosons
  which could result in a considerable changes of sgoldstino branching pattern~~\cite{Astapov:2014mea,Demidov:2016gmr}.
  %The third biggest decay channel is decay into a pair of $W$-bosons. Its branching ratio depends on sgoldsino mass and at least 3 times smaller than branching ratio of decay into a pair of gravitinos~\cite{Astapov:2014mea}
}.
Above expressions indicate that at fixed
$M_3$ very heavy sgoldstino decays mostly into gravitino pair, while
lighter sgoldstinos prefer to decay into gluons. For discussions of
sgoldstino branching patterns in more generic cases we refer an
interested reader to~\cite{Gorbunov:2002er,Asano:2017fcp,Astapov:2014mea}. 

\begin{figure}[!htb]
  \minipage{0.32\textwidth} 
  \includegraphics[width=\linewidth]{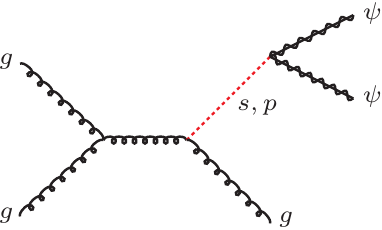}
  \endminipage\hfill
  \minipage{0.22\textwidth} 
  \includegraphics[width=\linewidth]{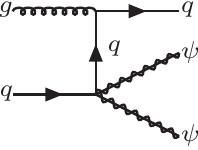}
  \endminipage\hfill
  \minipage{0.22\textwidth} 
  \includegraphics[width=\linewidth]{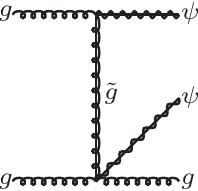}
  \endminipage\hfill
  \caption{\label{diag3} Examples of Feynman diagrams contributing to
    direct gravitino pair production with a jet.}
\end{figure}

Let us start with direct gravitino pair production with a single jet
\[
pp \to \psi \psi + {\rm jet}\,.
\]
On Fig.~\ref{diag3} we present examples of relevant Feynman diagrams
contributing to this subprocess.
Here we show a diagram with sgoldstino production, diagram with four-fermion
interaction and with gluino exchange in $t$-channel.
In the heavy sgoldstino limit, the cross section of this process behaves as
$1/F^4$. If sgoldstinos are light they can be produced on-shell
with subsequent decays into a pair of gravitinos.
If sgoldstino contribution is dominant, which happens for very light sgoldstinos, corresponding cross section scales as $1/F^2$.

\begin{figure}[!htb]
  \minipage{0.32\textwidth} 
  \includegraphics[width=\linewidth]{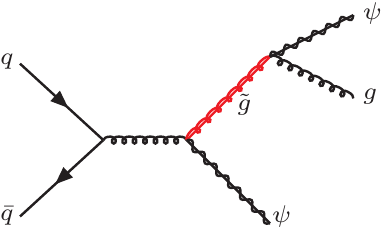}
  \endminipage\hfill
  \minipage{0.30\textwidth} 
  \includegraphics[width=\linewidth]{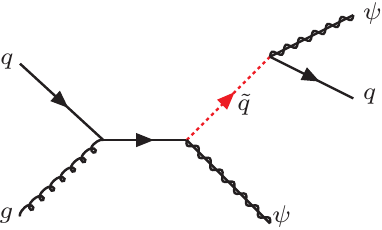}
  \endminipage\hfill
  \minipage{0.22\textwidth} 
  \includegraphics[width=\linewidth]{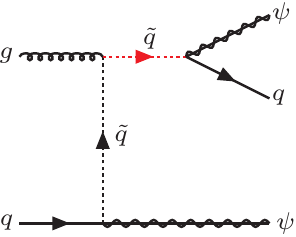}
  \endminipage\hfill
  \caption{\label{diag1} Examples of Feynman diagrams contributing to
    associated gravitino production with gluino or squark.}
\end{figure}

Next subprocess is associated gravitino production with gluino
or squark (see Fig.~\ref{diag1})
\[
pp \to \tilde{q} \psi, \tilde{g} \psi \to \psi \psi + {\rm jet}\,.
\]
In this case gluino or squark decay into gluino or quark, respectively,
with gravitino and hence we will have two gravitinos and jet in the final
state. The cross section behaves as $1/F^2$ at large values of SUSY
breaking scale.

\begin{figure}[!htb]
  \minipage{0.32\textwidth} 
  \includegraphics[width=\linewidth]{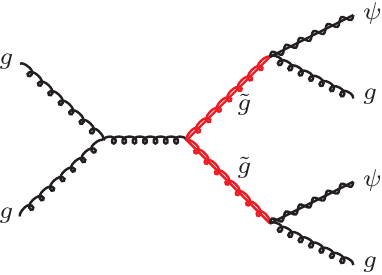}
  \endminipage\hfill
  \minipage{0.30\textwidth} 
  \includegraphics[width=\linewidth]{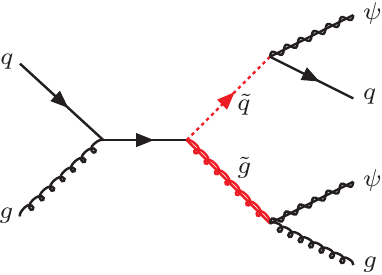}
  \endminipage\hfill
  \minipage{0.22\textwidth} 
  \includegraphics[width=\linewidth]{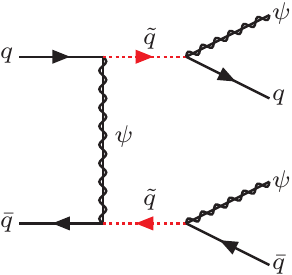}
  \endminipage\hfill
  \caption{\label{diag2} Examples of Feynman diagrams contributing to
    SUSY QCD pair (squark-squark, gluino-gluino and squark-gluino)
    production.}
\end{figure}

At last one has squark-squark, gluino-gluino and
squark-gluino production (or SUSY QCD pair production) with their subsequent decays into gravitino
and gluons or quarks (see Fig.~\ref{diag2})
\[
pp \to \tilde{q}\tilde{q},\tilde{q}\tilde{g},\tilde{g}\tilde{g} \to \psi \psi + 2~{\rm jets}\,.
\]
The final state of these subprocesses contains in
fact two jets but the event selection procedure used by ATLAS and CMS
experiments for monojet searches actually admit more than one
jet in a final state. It appears that SUSY QCD pair production gives quite significant
contribution, thus it is important to take these processes into 
account especially at large values of SUSY breaking scale where they
are dominant and their cross section does not depend on $\sqrt{F}$.

In the case of TeV scale sgoldstino other types of subprocesses
contributing to jet-plus-missing-energy signature at tree level are
possible. As it follows from \eqref{mdl:lgr5} the interaction
lagrangian contains higher dimensional operators of the following
types  
\begin{equation}
  \frac{m_{\rm soft}^2}{F^2}\tilde{q}\partial_\mu\phi\, \bar{q}\gamma^\mu\psi,\;
  \frac{m_{\rm soft}^2}{F^2}D_\mu\tilde{q}\phi\, \bar{q}\gamma^\mu\psi, \; {\rm etc}.
\end{equation}
They would generate subprocesses like $qg\to\tilde{q}\phi\psi$, which
after subsequent decays $\tilde{q}\to q\psi$ and $\phi\to\psi\psi$
would result in missing energy signature. However, we expect that
cross section of these subprocesses
will be suppressed with respect to SUSY QCD  pair production and to
gravitino associated production with squark(gluino) by additional
power of $m_{\rm soft}/F$ suppression \footnote{We remind that self-consistency of the effective theory we consider here implies that inequality $m_{\rm soft}<\sqrt{F}$ must be fulfiled.}. It would also endure phase space suppression due to production of two
heavy particles (sgoldstino and squark). In what follows we neglect
contribution of such subprocesses.

We implemented model~\eqref{mdl:lgr6}--\eqref{mdl:lgr7} into \texttt{MadGraph}~\cite{Alwall:2014hca} and calculated cross section for the processes in question at the leading order applying all the relevant cuts from CMS monojet analysis with run-I data~\cite{Khachatryan:2014rra}.
This CMS study selects events with at most two jets in the final state where the primary jet
($j_{1}$) should have transverse momentum $p_T(j_1) > 110~{\rm GeV}$
and pseudorapidity $\vert \eta (j_1) \vert~<~2.4$. The secondary jet
$(j_2)$ is also allowed if $p_T(j_2) > 30~{\rm GeV}$, $\vert \eta
(j_2) \vert < 4.5$ and difference in azimuthal angle $\vert \Delta
\phi (j_1,j_2) \vert~<~2.5$. For cross check we verify obtained cross sections using \texttt{CalcHEP}~\cite{Belyaev:2012qa}.

In general the model in question contains a lot of free parameters:
\[
M_{3},~M_{\tilde{q}_{L},ij},~M_{\tilde{u}_{R},ij},~M_{\tilde{d}_{R},ij},~A_{d,ij},~A_{u,ij},~\sqrt{F},~m_{s},~m_{p},
\]
were $i,j$ is a family index. We however consider simplified class of models by making several assumptions. Firstly, we assume no mixing in squark sector, which means that $M_{\tilde{q}_{L},ii},~M_{\tilde{u}_{R},ii},~M_{\tilde{d}_{R},ii}$ are physical squark masses. Next, we assume equal masses of gluinos and all squarks
\[
M_{\tilde{q}_{L},ii} = M_{\tilde{u}_{R},ii} = M_{\tilde{d}_{R},ii} \equiv M_{\tilde{q}} = M_{3}.
\]
Equality of gluino and squark masses implies that these particles decay mostly into gravitinos and gluino and quark.
Furthermore, we assume that scalar and pseudoscalar sgoldstinos are degenerate in mass, $m_s=m_p$ and take into account only their decays into pair of gluons and gravitinos which are dominant for TeV-scale gluinos and sgoldstinos. For calculations of cross sections we use CTEQ6L1 PDFs~\cite{Pumplin:2002vw} with relevant
values of renormalization and factorization scales for different
subprocesses contributing to the signal.
Namely, for the process of direct gravitino pair production we take
$\mu_R=\mu_F = \frac{1}{2}(p_T^j + m_T^{\psi\psi})$, where
$m_T^{\psi\psi} = \sqrt{{m^{\psi\psi}}^2 + {p_T^j}^2}$ and
$m^{\psi\psi}$ is invariant mass of pair of gravitinos.
For gluino/squark associated production with gravitino we take
$\mu_{R} = \mu_{F} = 0.5 M_{\tilde{q}} = 0.5 M_{3}$ while for
QCD pair production  $\mu_{R} = \mu_{F} = M_{\tilde{q}} = M_{3}$.
Unstable particles in the intermediate states are taken into
account in narrow width approximation (NWA) which implies that the results are valid when
\[
m_{s} \gg \frac{m_{s(p)}^{5}}{32 \pi F^2} + \frac{M_3^2 m_{s(p)}^{3}}{4 \pi F^2}
\]
Unitarity arguments also constrain the ratio between sgoldstino masses and $\sqrt{F}$. Namely, application of optical theorem to process $\psi \psi \to \psi \psi$ yields the following constraint on sgoldstino masses~\cite{Brignole:1996fn}
\[
m_{s}^{2} + m_{p}^{2} < \sqrt{48\pi} F,
\]
which in our case of degenerate sgoldstino masses can be rewritten as \linebreak $m_{s}(m_{p}) < \sqrt{2 \sqrt{3\pi}} \sqrt{F}\approx 2.5\sqrt{F}$.

CMS analysis considers a number of requirements for the amount of $\bcancel{p_T}^{\rm min}$ in the following set:
\[
\bcancel{p_T^{\rm min}} \in \{ 250, 300, 350, 400, 450, 500, 550 \}~{\rm GeV}
\]
We checked that for the parameter space we use here the strongest
bound comes from the requirement $\bcancel{p_T^{\rm min}}=450~{\rm GeV}$. In
this case the upper limit on the visible cross section times
acceptance times efficiency for non-SM production of events is about
7.8~fb. We do not take into account detector effects; corresponding
efficiencies are expected to be close to unity for the selected
events, see~\cite{Khachatryan:2014rra}.
In~\cite{Maltoni:2015twa} an additional cut $p_T(j_2)<150$~GeV was
imposed to decrease contribution of SUSY QCD subprocesses. It was
motivated by the fact that cross section of these subprocesses is not
sensitive to $\sqrt{F}$, i.e. to gravitino mass. Still as we will see
below large contribution of SUSY QCD subprocesses to the signal allow
us to exclude some models with relatively light masses of
superpartners within framework of low scale supersymmetry breaking
using missing transverse energy signature. For this reason we do not
introduce any additional cuts in the phase space.

Let us briefly comment on NLO corrections. It is known that for squark and gluino productions they are
large~\cite{Beenakker:1996ch,Beenakker:2010nq,GoncalvesNetto:2012yt}. For
instance, corresponding K-factor reaches value of 3 for gluino-gluino 
production~\cite{Beenakker:1996ch,Beenakker:2010nq,GoncalvesNetto:2012yt}. However,
our whole theory which includes gravitinos and sgoldstinos is 
effective and nonrenormalizable. To perform self-consistent NLO
analysis one should 
have knowledge of microscopic theory behind it. This is beyond
the scope of our work. Given the fact the NLO corrections typically
increase the production cross sections we expect that our bounds will
be even stronger if we include them into our analysis. Also it has
been shown in previous work~\cite{Maltoni:2015twa}  that effects of
showering on jet transverse missing energy distribution are important
but rather mild. We will neglect them here in view of large expected
NLO corrections. 

\begin{figure}[!htb]
  \begin{tabular}{ll}
    \includegraphics[width=0.485\columnwidth]{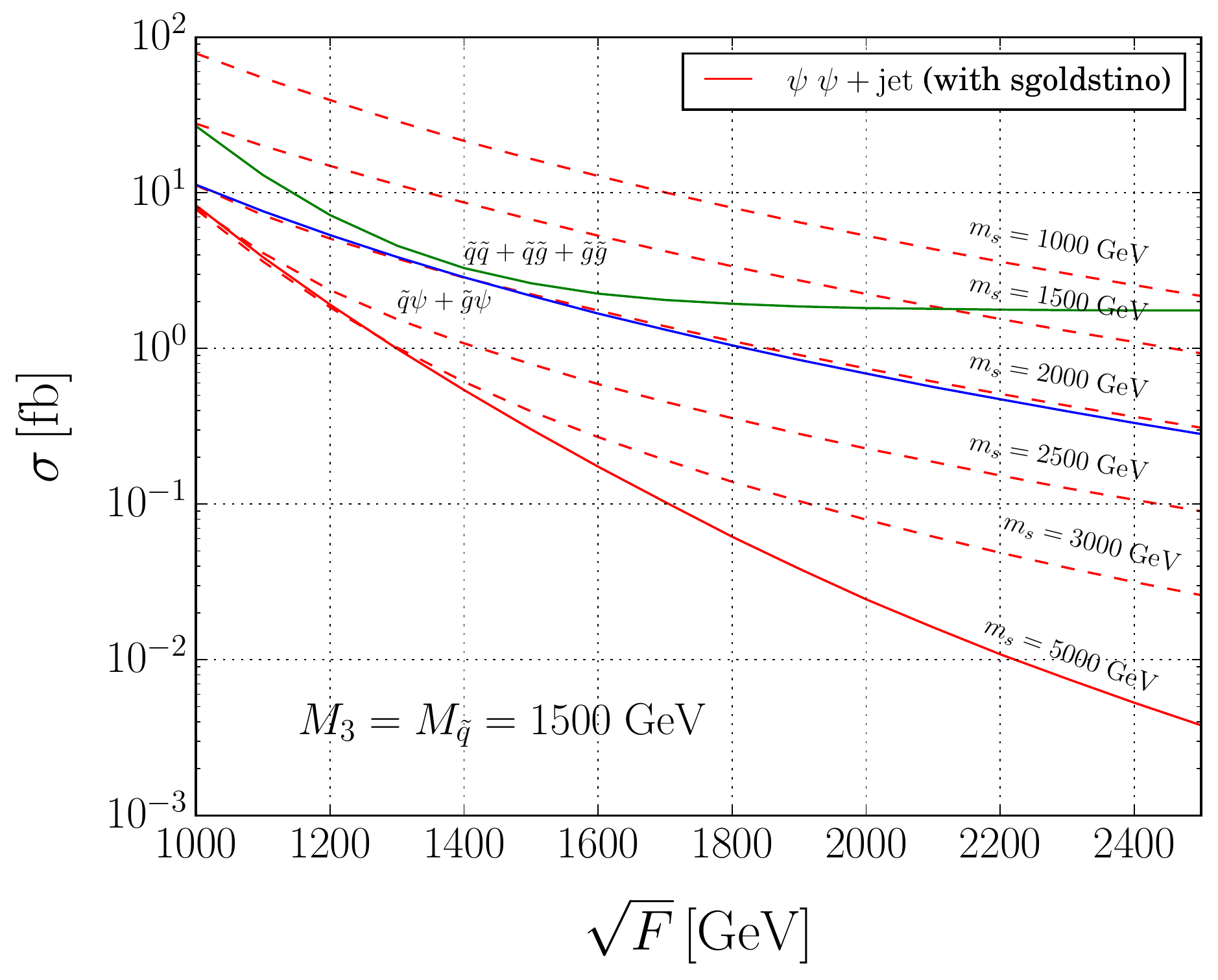} &
    \includegraphics[width=0.485\columnwidth]{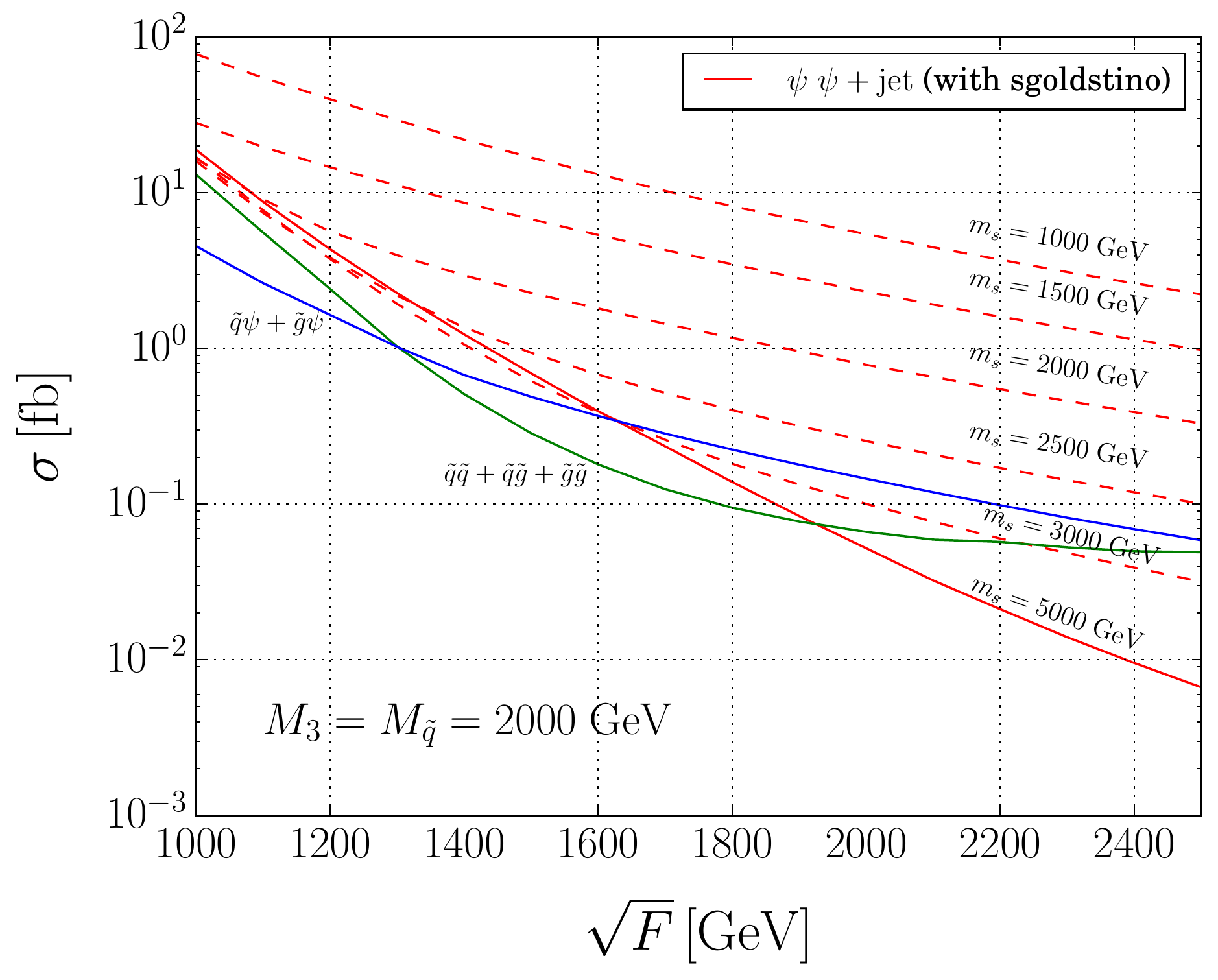}
  \end{tabular}
  \caption{\label{fig1} Cross sections of relevant subprocesses
    contributing to $pp \to \psi\psi + {\rm jet} \to \bcancel{p_{T}} + {\rm jet}$ as functions of supersymmetry breaking scale
    for $M_3=M_{\tilde{q}}=1.5$~TeV (left panel) and
    $M_3=M_{\tilde{q}}=2$~TeV (right panel). }
\end{figure}

On Fig.~\ref{fig1}
we show cross sections of relevant subprocesses as functions of
supersymmetry breaking scale. On the left panel masses of
superpartners are equal to 1.5~TeV and one can see that for heavy 
sgoldstinos the superpartner production and associated production are
the dominant processes. However for sgoldstinos  with masses of
several TeVs corresponding cross sections can be comparable or even
larger. Similar behaviour can be observed on the right panel on the
Figure (where the masses of superpartners are taken to be 2~TeV) but 
with different hierarchy between different subprocesses. One can see
that steep slope of gravitino pair production in the heavy sgoldstino
mass limit becomes more flat with light sgoldstino which respects
changing of cross section scaling from $1/F^4$ to $1/F^2$ with the
increase of light sgoldstino contribution.

\begin{figure}[!htb]
  \begin{tabular}{ll}
    \includegraphics[width=0.485\columnwidth]{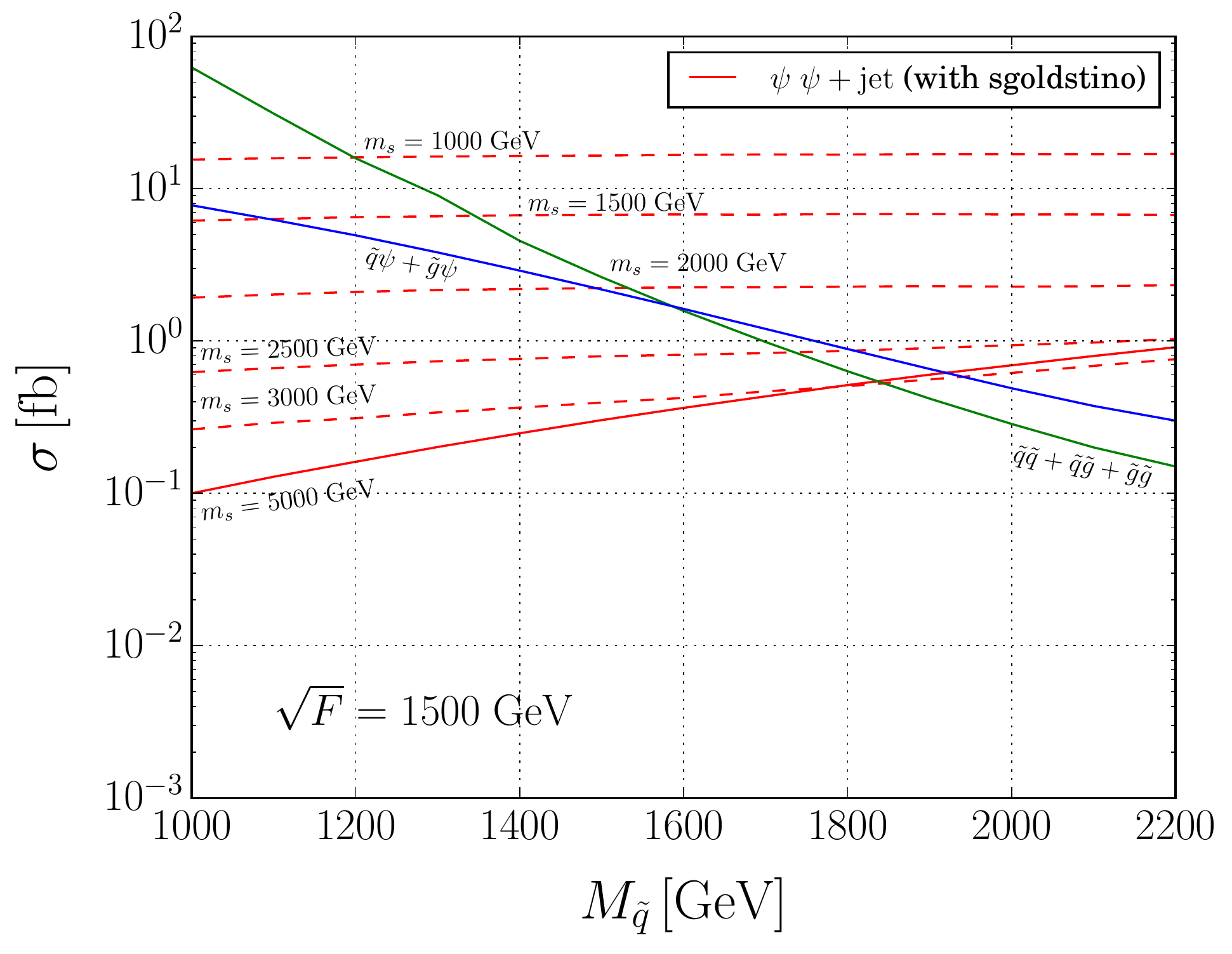} &
    \includegraphics[width=0.485\columnwidth]{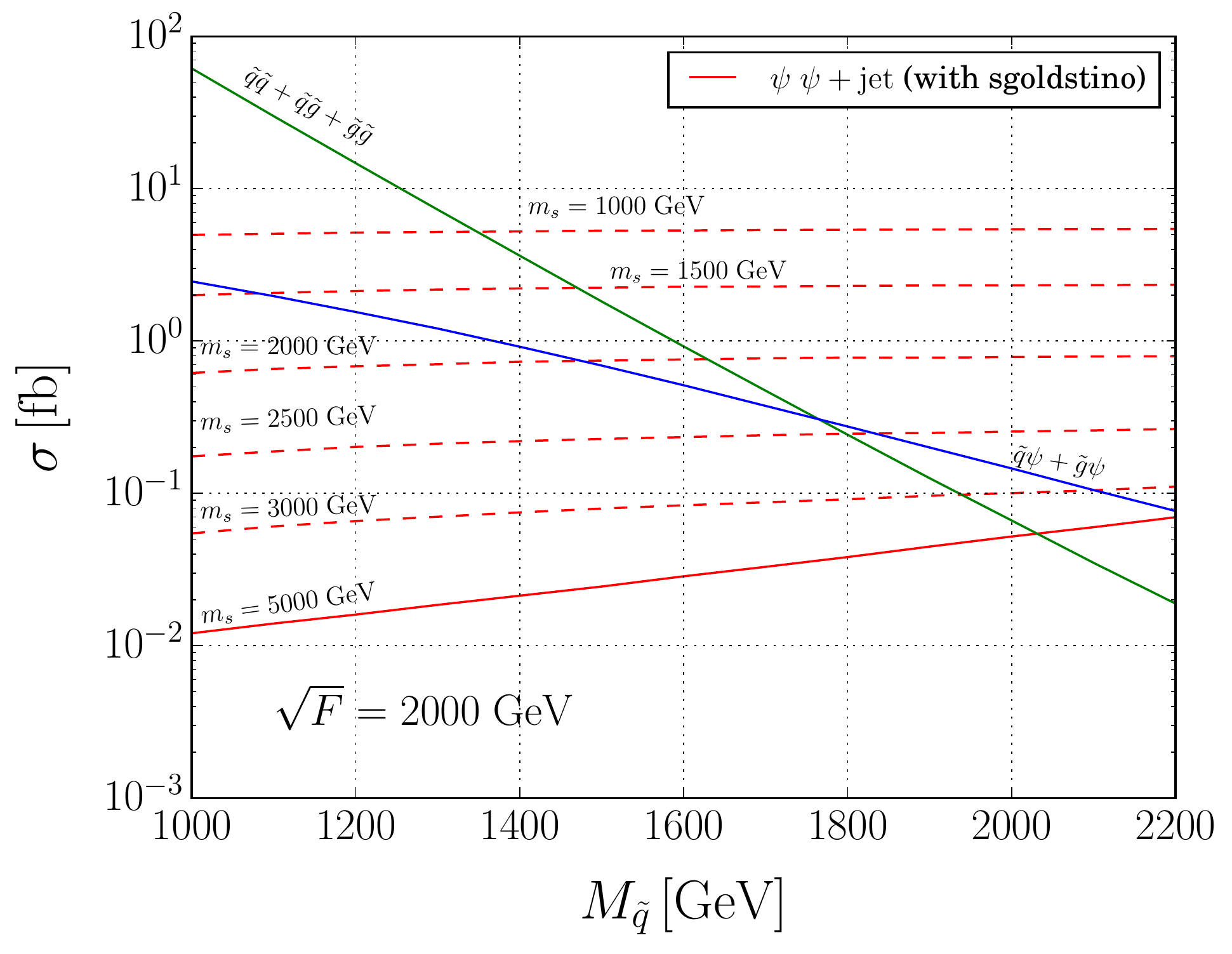}
  \end{tabular}
  \caption{\label{fig2} Cross sections of relevant subprocesses
    contributing to $pp \to \psi\psi + {\rm jet} \to \bcancel{p_{T}} + {\rm jet}$ as functions of common mass of superpartners
    for $\sqrt{F}=1.5$~TeV (left panel) and $\sqrt{F}=2$~TeV (right
    panel). }
\end{figure}

On Fig.~\ref{fig2}
we show the same cross sections but as functions of common
mass of superpartners at different values of SUSY breaking
scale. Obviously, production cross section of superpartners decreases
with increase of their masses. On the contrary, direct gravitino pair
production increases and stays constant at large masses of
squarks. Let us note that the contribution of light sgoldstino is
prominent as compared with heavy sgoldstino limit and can increase
cross section of the corresponding subprocess by almost 3 orders of
magnitude.

\begin{figure}[!htb]
\begin{center}
  \includegraphics[width=0.9\columnwidth]{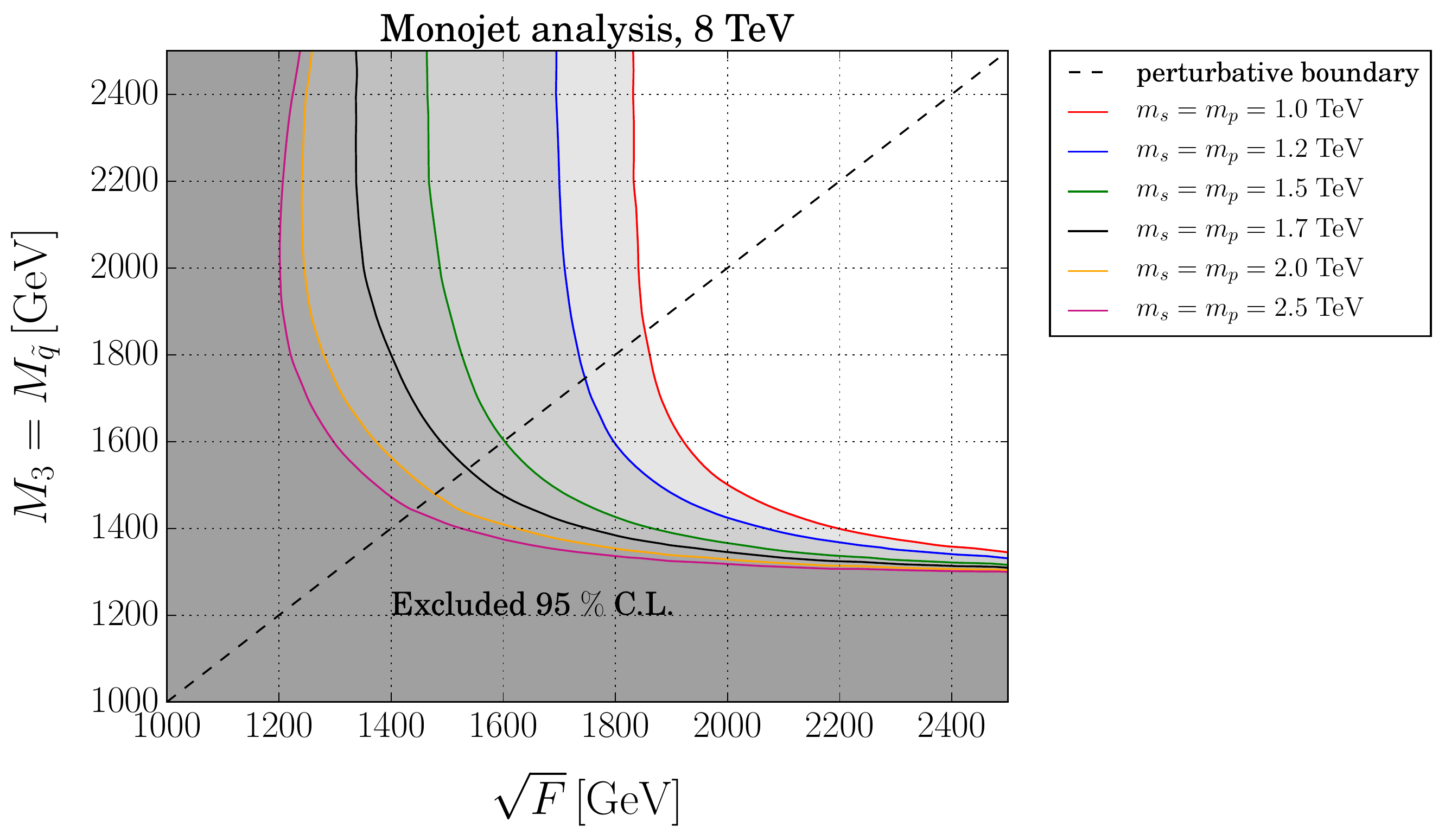}
  \end{center}
  \caption{\label{fig3} Exclusion plot of $\sqrt{F}$ vs
    $M_3=M_{\tilde{q}}$     for different masses of sgoldstino.}  
\end{figure}

On Fig.~\ref{fig3}
we show exclusion plots of SUSY breaking scale $\sqrt{F}$ versus
common mass of superpartners $M_3=M_{\tilde{q}}$ at fixed values of 
sgoldstino masses. One can see that light sgoldstinos can change the
bounds on $\sqrt{F}$ (and as consequently on the gravitino mass $m_{3/2}$)
considerably. Flattening of these lines at large values of SUSY 
breaking scale is due to saturation of the total cross section by by
pair production of superpartners, squarks and gluinos, in this
limit. However, one should remember that at very large value of $\sqrt{F}$ squarks/gluinos cease to decay
dominantly into quark/gluino and gravitinos (which is one of the assumptions of the present analysis) and the bounds from the monojet searches should be weakening.
On the opposite end, at large $M_3=M_{\tilde{q}}$ the
contour becomes almost vertical because in this parameter region cross section is dominated by
direct gravitino pair production in association with jet. In between
all the subprocesses are of the same relevance. The dashed line here
is the boundary of applicability of perturbation theory for our
model $m_{\rm soft}<\sqrt{F}$. Above this line the theory is in the strong coupling regime. 

\begin{figure}[!htb]
  \begin{center}
    \includegraphics[width=0.9\columnwidth]{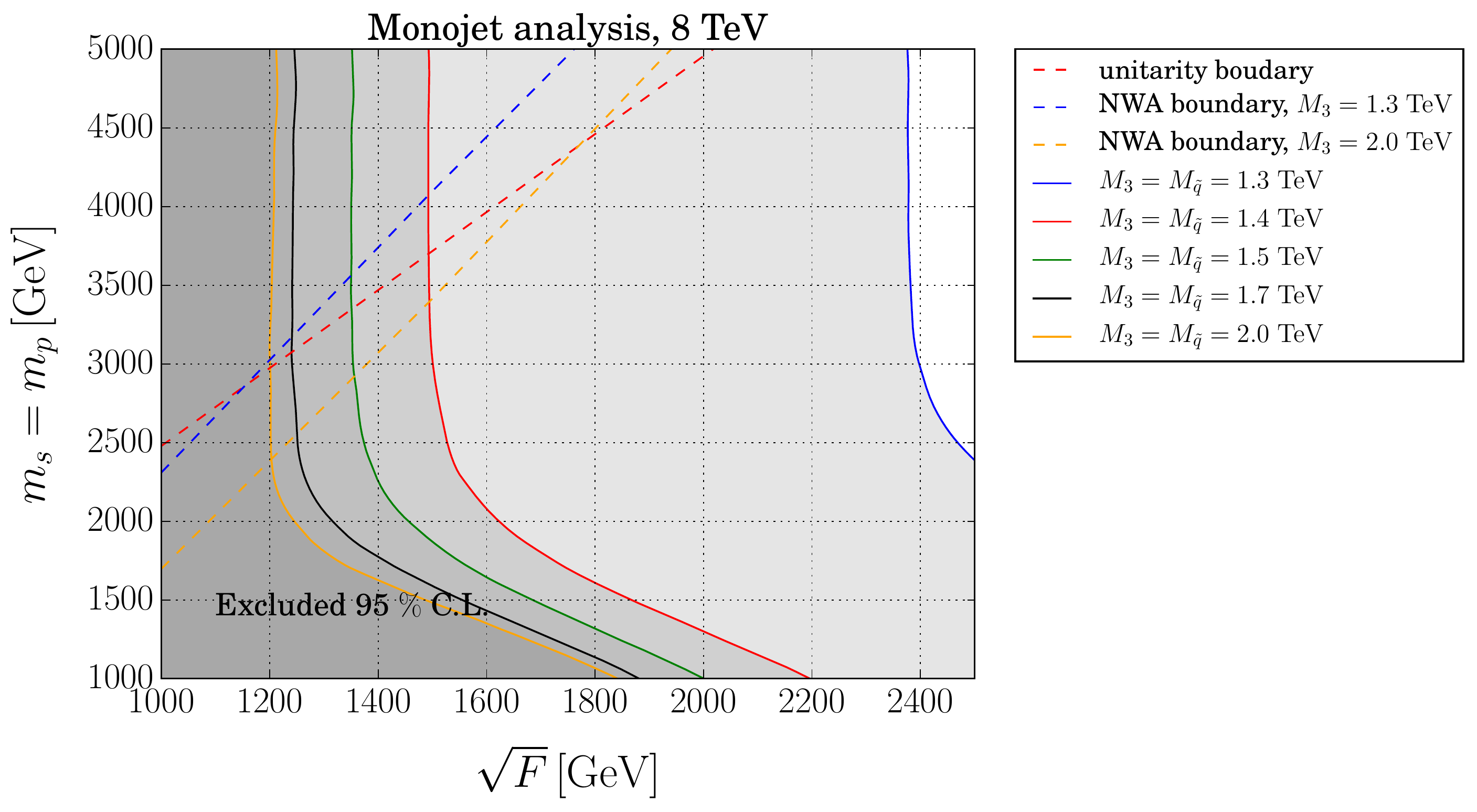}
    \end{center}
  \caption{\label{fig4}Exclusion plot of $\sqrt{F}$ vs $m_s=m_p$ for
    different scale of masses of superpartners. Above the dashed lines one of the corresponding conditions is violated: unitarity, condition for applicability of NWA. Here we depicted only lines which correspond to $M_{3} = 1.3~{\rm TeV}$ (blue dashed) and $M_{3} = 2.0~{\rm TeV}$ (orange dashed). All other lines which correspond to all intermediate scales lie between them.} 
\end{figure}

On Fig.~\ref{fig4}
we present another exclusion plot but in different coordinates. Here
we fix the mass scale of superpartners and vary common masses of
sgoldstinos and SUSY breaking scale. One can see that sgoldstino
contribution is the most important in the range of its mass from 1~TeV
to 2.5~TeV where the bounds on SUSY breaking scale are strengthen by
factor up to about 1.5 in comparison with those obtained in the heavy
sgoldstino limit. This corresponds to a factor of 2 in limit on gravitino mass $m_{3/2}$. Let us note that for $m_s=m_p \gtrsim \sqrt{F}$
narrow width approximation which is used in this work is not applicable~\footnote{Unitarity condition is violated as well in this case.}. But in this limit sgoldstino contribution is suppressed and we do not expect considerable changes in our results.

\section{Direct sgoldstino production and dijet signature}

Light sgoldstinos can be produced directly in $pp$ collisions mainly in
gluon-gluon fusion~\cite{Perazzi:2000ty} and after their decay into
pair of gluons they can be observed as narrow dijet resonances. It is
interesting to compare the bounds from monojet analysis with the bounds
which can be obtained from the the LHC searches of dijet resonance.
For comparison we use both ATLAS~\cite{Aad:2014aqa} and
CMS~\cite{Khachatryan:2015sja} dijet analyses with the data obtained
at $\sqrt{s}=8$~TeV. Although new results of both experiments
at $\sqrt{s}=13$~TeV are already
available~\cite{Aaboud:2017yvp,Sirunyan:2016iap} we still use run-I
data of the  same statistics as we used for monojet analysis in the previous Section. 

We use \texttt{MadGraph}~\cite{Alwall:2014hca} to calculate leading order cross
section of the process $pp\to s(p)\to 2~{\rm jets}$ at the partonic level.
To apply ATLAS upper limits on dijet cross section we impose the
following cuts: $\vert \eta(j_{1,2}))\vert < 2.8$, $p_T(j_{1,2}) >
50$~GeV with $\frac{1}{2}\vert \eta(j_1)-\eta(j_2)\vert < 0.6$ and
dijet invariant mass $m_{jj}>250$~GeV. For the case of the CMS results
we use the following set of cuts: $\vert \eta \vert < 2.5$ and scalar
sum of gluon $p_T$, $H_T>150$~GeV with either $H_T>650$~GeV or
$m_{jj}>750$~GeV with $\vert  \eta(j_1)-\eta(j_2) \vert < 1.5$. In both cases additional requirement $m_{jj}>890$~GeV
should be fulfilled. We do not include detector effects and showering in
view of unknown NLO corrections to the cross section.

\begin{figure}[!htb]
  \begin{tabular}{ll}
    \includegraphics[width=0.45\columnwidth]{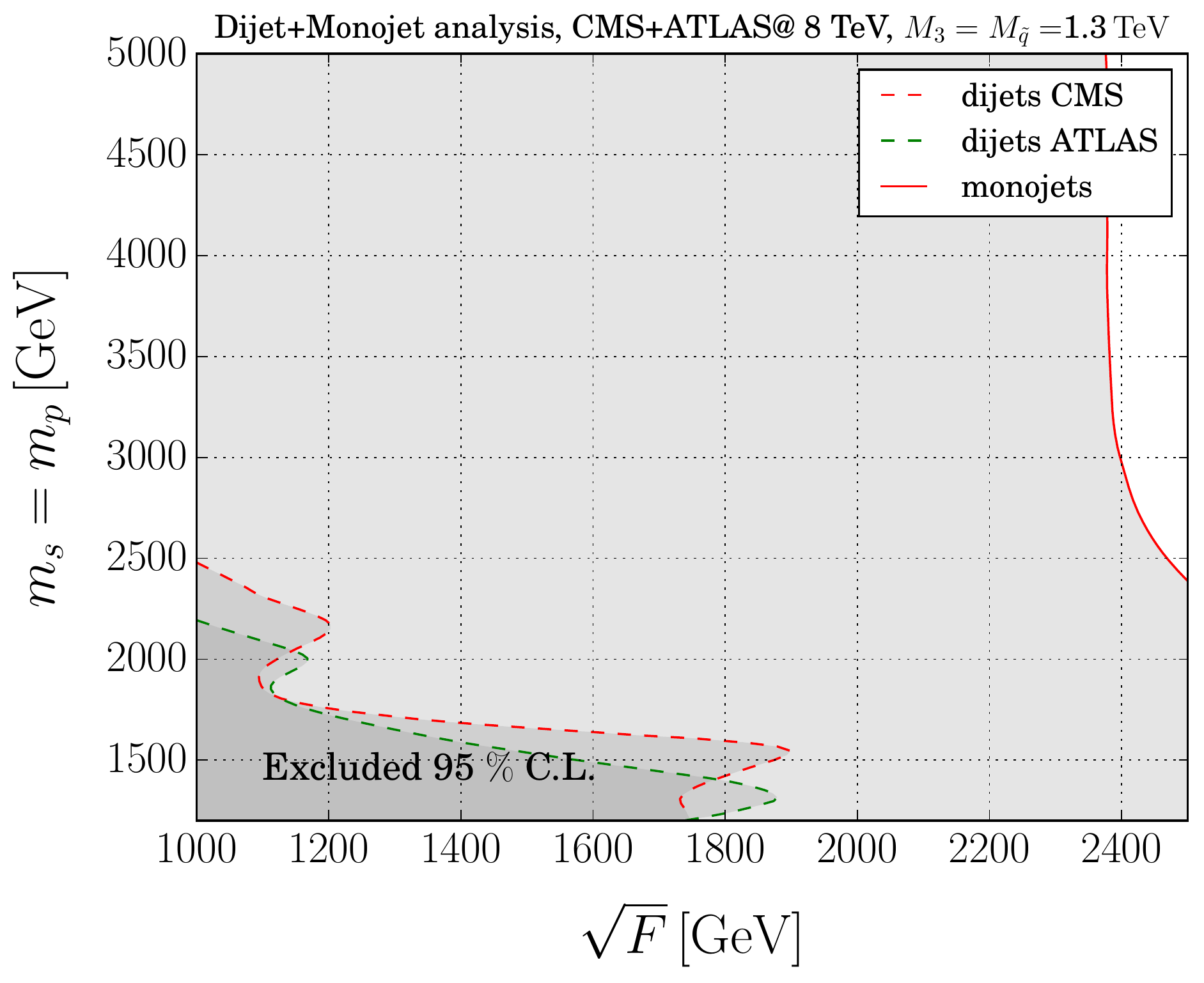} &
    \includegraphics[width=0.45\columnwidth]{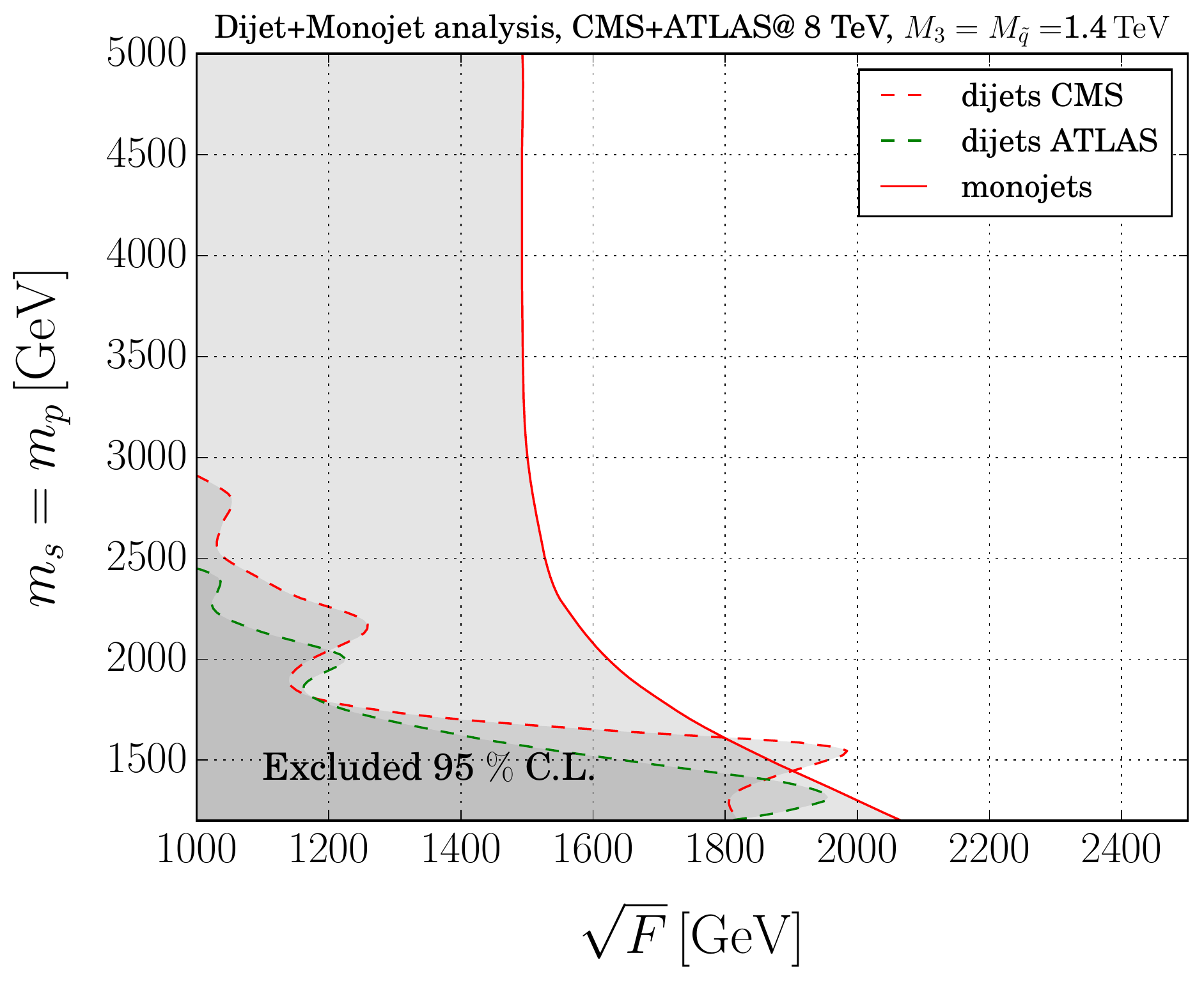} \\
    \includegraphics[width=0.45\columnwidth]{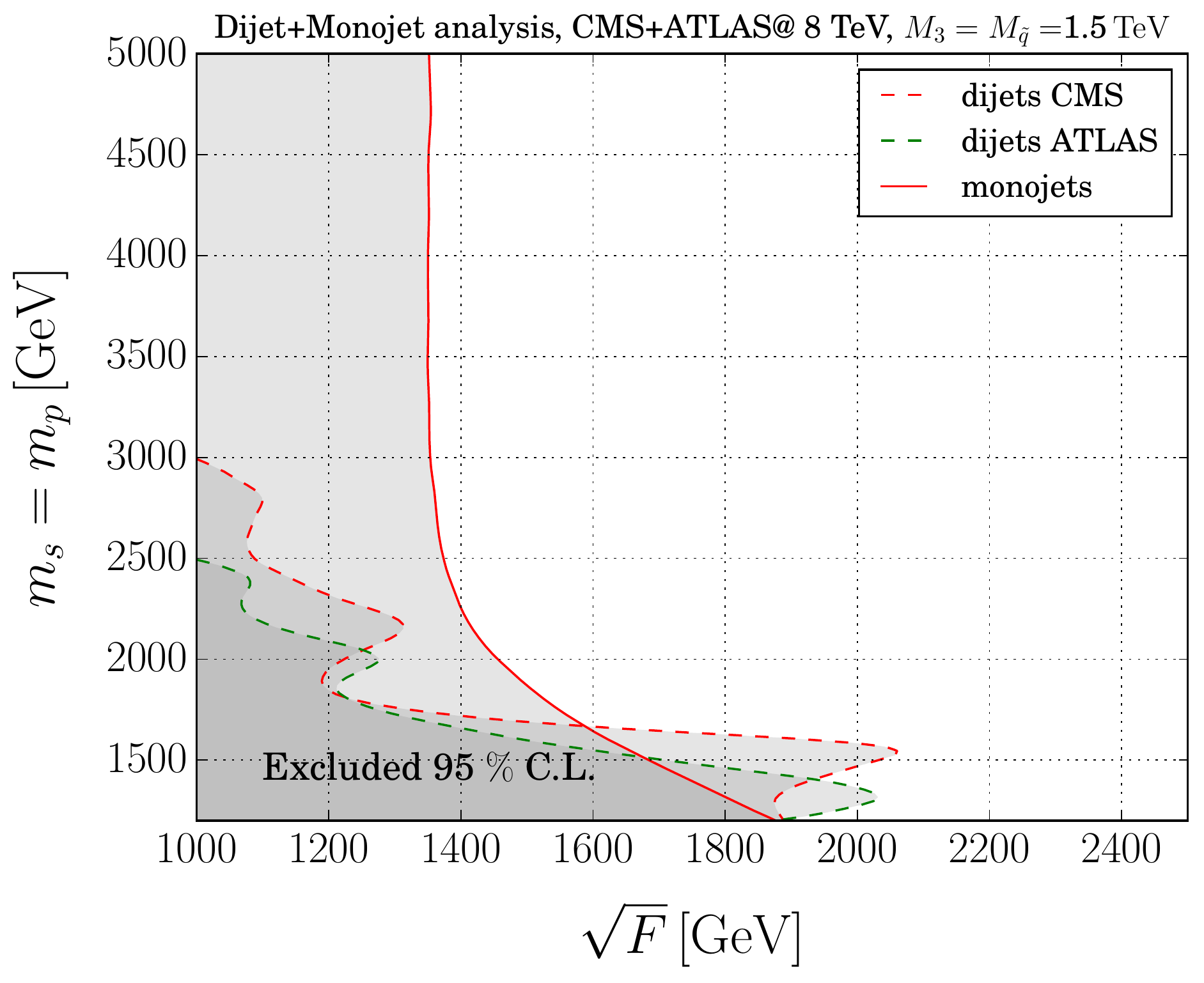} &
    \includegraphics[width=0.45\columnwidth]{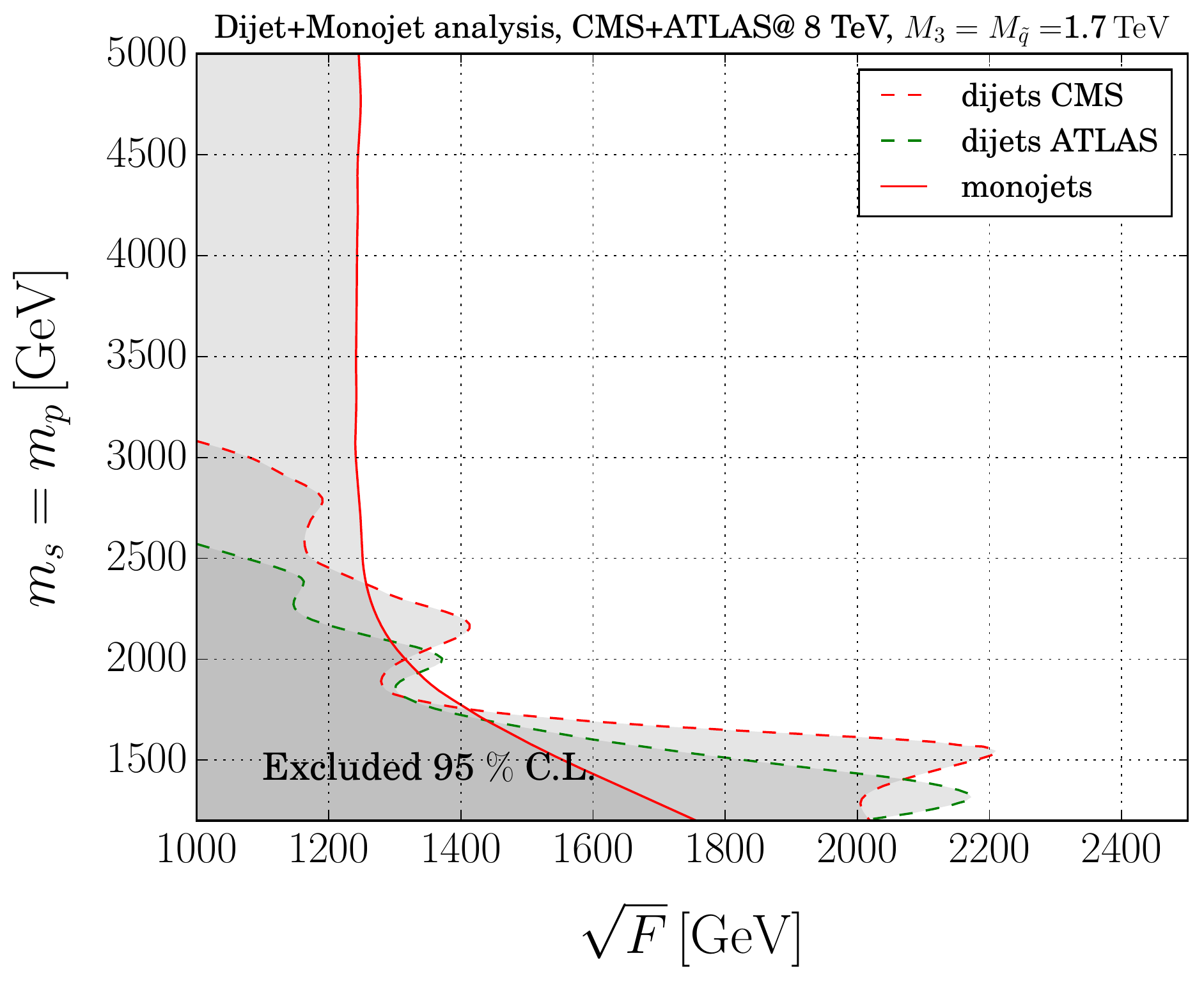}
  \end{tabular}
  \caption{\label{compare1} Exclusion plots of $\sqrt{F}$ vs
    $m_s=m_p$, comparison of bounds obtained from dijet resonances and
    missing energy searches for different values of
    $M_3=M_{\tilde{q}}$: 1.3~TeV (upper left), 1.4~TeV (upper right),
    1.5~TeV (lower left), 1.7~TeV (lower right).} 
\end{figure}

To find excluded models we've compared calculated cross sections with the
experimental upper 95~$\%~{\rm C.L.}$ limits on the dijet cross sections obtained by
ATLAS~\cite{Aad:2014aqa} and CMS~\cite{Khachatryan:2015sja}. 
On Fig.~\ref{compare1}
we present comparison of exclusion plots of SUSY breaking scale
$\sqrt{F}$ vs sgoldstino mass $m_s=m_p$ for different values of
superpartners masses $M_3=M_{\tilde{q}}$ obtained from dijet and
monojet searches. One can see that for relatively small masses of
squarks and gluinos monojet analysis limits this models considerably
stronger than dijets. In this case the monojet cross section due to
contribution of these superpartners. With the increase of masses of
superpartners direct production of sgoldstinos with their subsequent
decays into pair of gluons becomes more constraining. But for heavy
sgoldstinos it weakens again for two reasons: on the one hand direct
production of sgoldstinos becomes suppressed by its mass and on the
other in this case sgoldstinos decay dominantly a into pair of
gravitinos. In this way searches for monojets and dijets are
complimentary to each other.

\begin{figure}[!htb]
  \begin{tabular}{ll}
    \includegraphics[width=0.45\columnwidth]{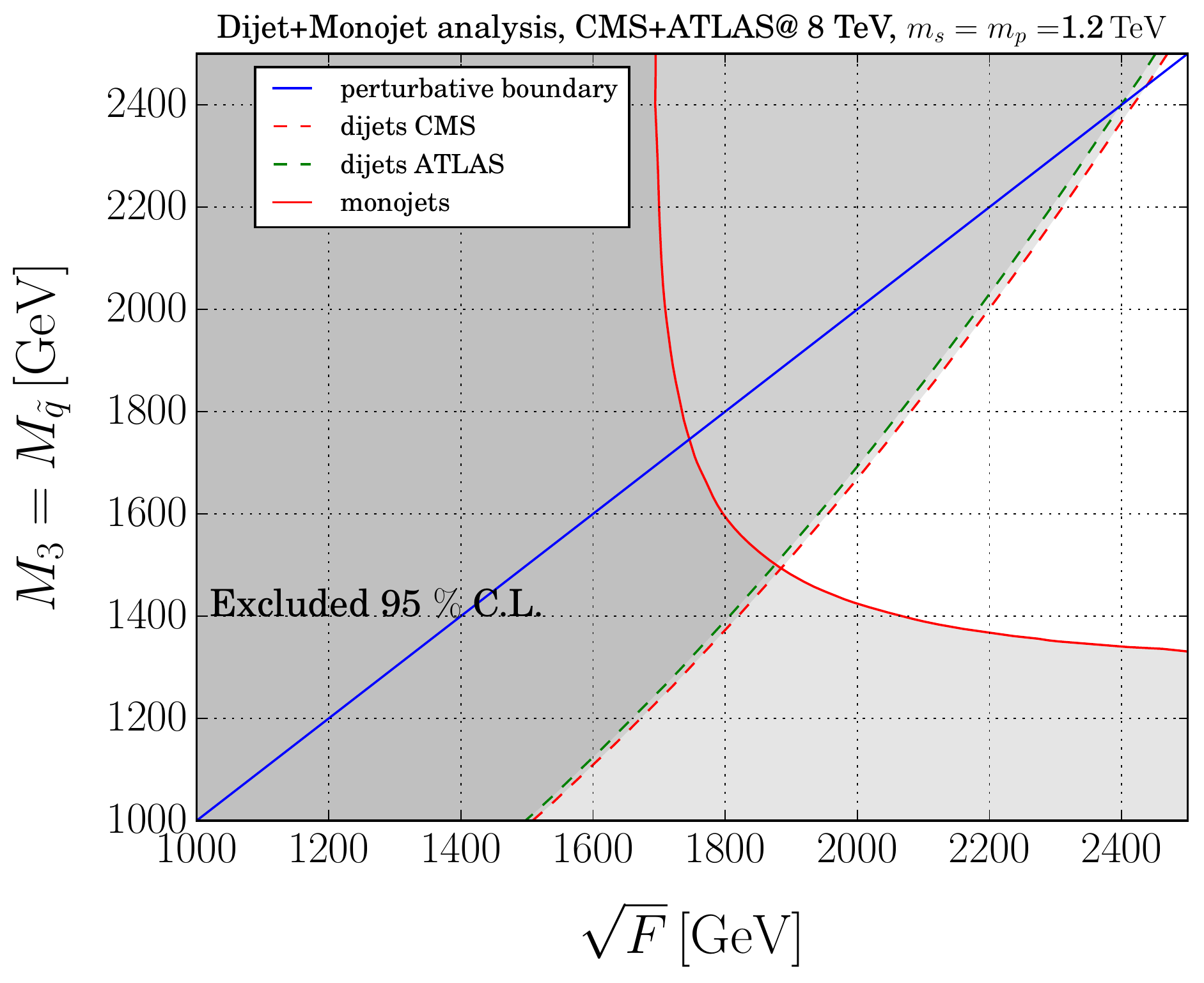} &
    \includegraphics[width=0.45\columnwidth]{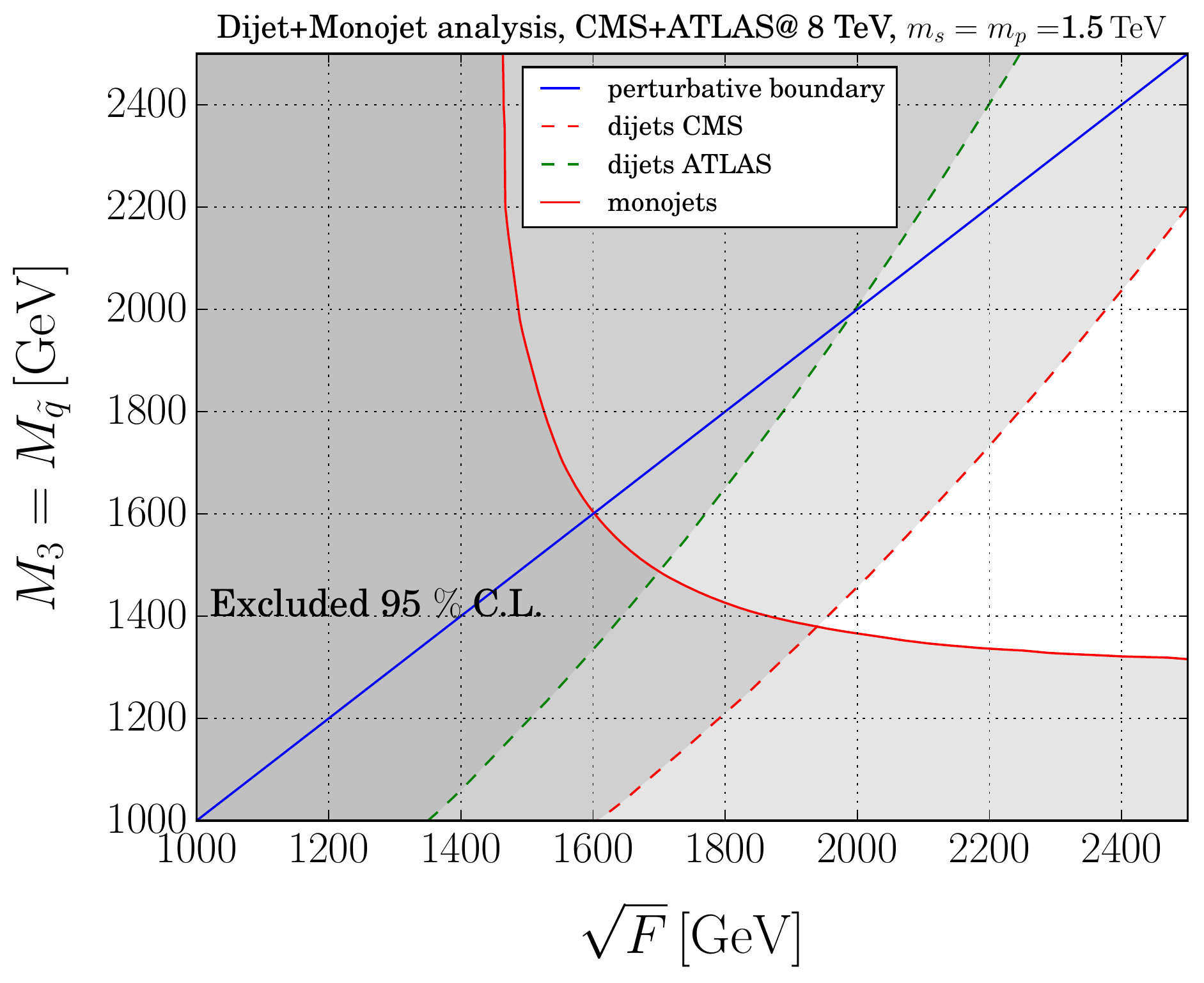} \\
    \includegraphics[width=0.45\columnwidth]{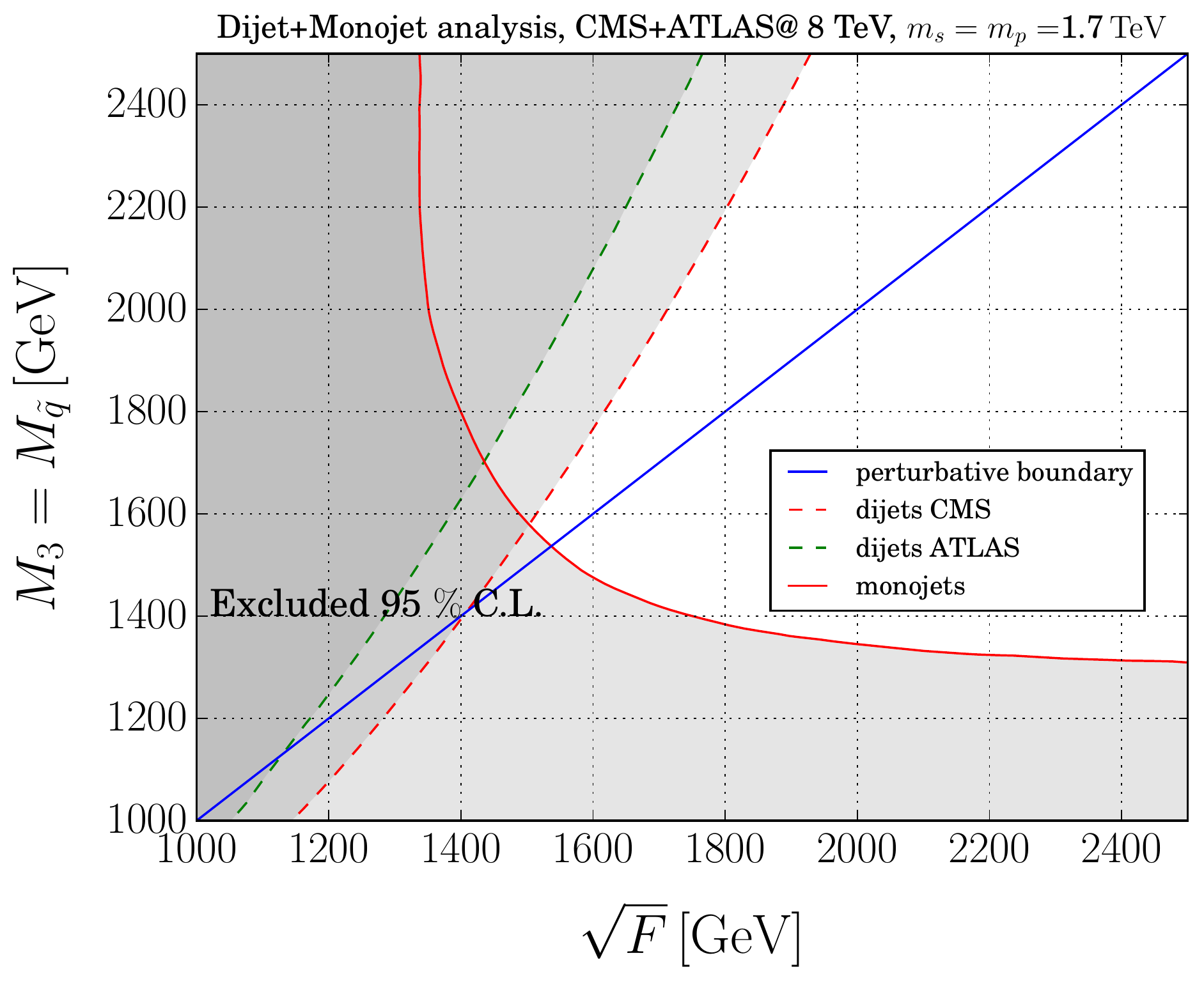} &
    \includegraphics[width=0.45\columnwidth]{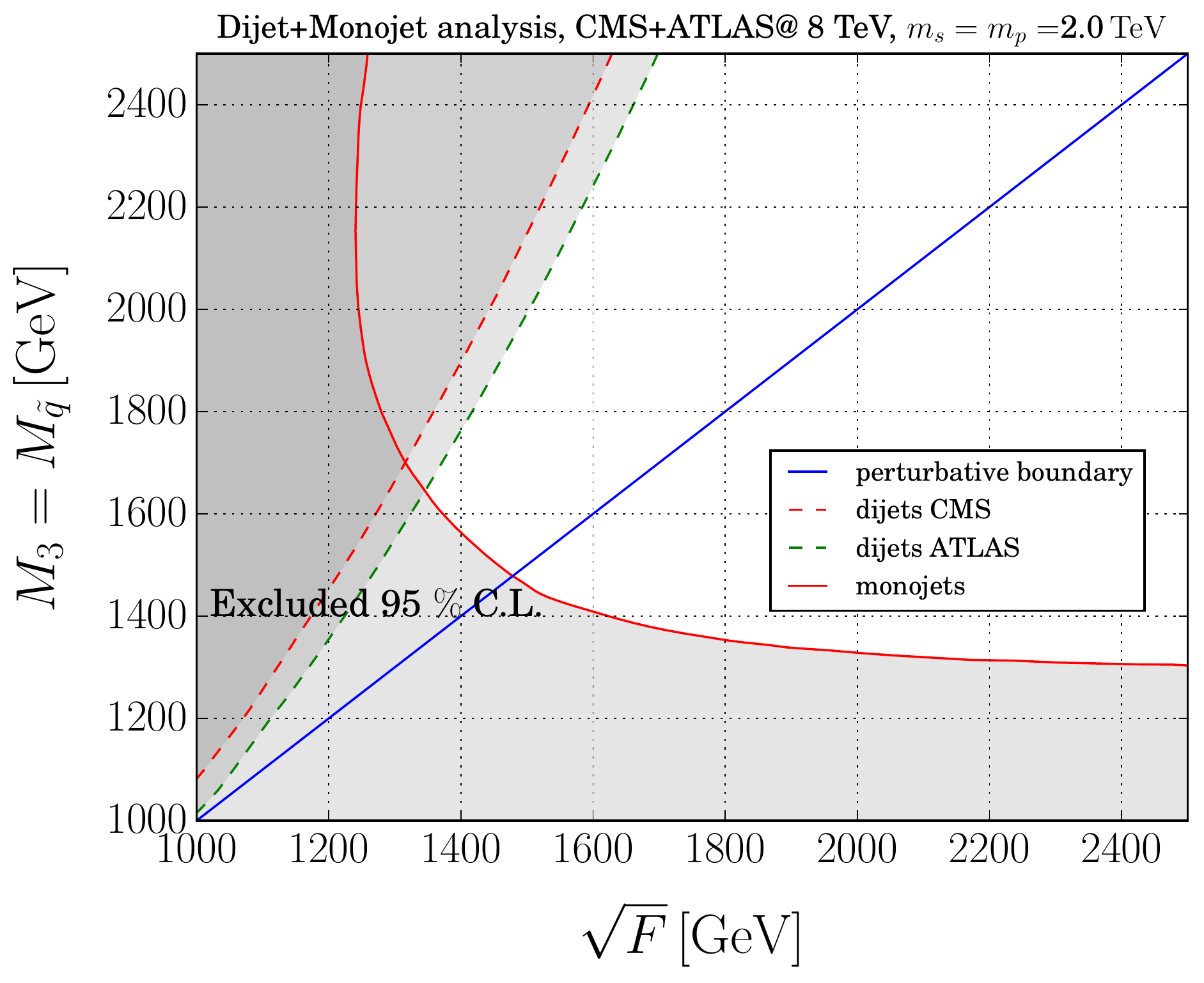}
  \end{tabular}
  \caption{\label{compare2} Exclusion plots of $\sqrt{F}$ vs
    $M_3=M_{\tilde{q}}$, comparison of bounds obtained from dijet resonances and
    jet~$+$~missing energy searches for different values of
    $m_s=m_p$: 1.2~TeV (upper left), 1.5~TeV (upper right),
    1.7~TeV (lower left), 2.0~TeV (lower right)} 
\end{figure}

On Fig.~\ref{compare2}
we present similar exclusion plots but in different coordinates:
supersymmetry breaking scale $\sqrt{F}$ vs $M_3=M_{\tilde{q}}$ for
different selected values of sgoldstino masses. For instance, in the
upper left panel sgoldstino mass is equal to 1.2~TeV and here we see
again that at small masses of superpartners the constraints from
jet~$+$~missing transverse energy searches are stronger than from dijets,
while for heavy superpartners it is vice versa. For sgoldstinos with mass more than $\sim 1.7~{\rm TeV}$ the bounds from dijets searchs, being formally more
stringent for large masses of superpartners, actually lie in a strong
coupling regime of the theory $m_{\rm soft}>\sqrt{F}$  and hence in this case constraints from monojet
searches are the most relevant.

\section{Conclusions}
\label{conclusions}

To summarize in this paper we show that in models with low scale
supersymmetry breaking contribution of light sgoldstino to gravitino
pair production can be considerable, although its actual size depends
on the mass of sgoldstino. We calculate leading order cross sections of
the processes contributing to jet and missing energy signal. We obtain bounds on the parameter space of the model within
a simplified set of parameters using results of the CMS run-I searches
for jet and missing transverse energy signature at $\sqrt{s}=8$~TeV.
We found that the bounds on gravitino mass in the case of light
sgoldstinos can be stronger by factor of 2 as compared to those in the
heavy sgoldstino limit. We compare the bounds from monojet searches
with those from dijets from ATLAS and CMS data of the same collision energy and
statistics.  We found them complimentary in different regions of parameter space. Our final results are presented in Figs.~\ref{compare1} and~\ref{compare2}.
 For instance, for common mass of superpartners 1.5~TeV the bound on supersymmetry breaking scale varies from
  about 1.35~TeV for heavy sgoldstinos to about 2~TeV for sgoldstinos with mass about 1.5~TeV. The respective bounds on $m_{3/2}$ in this case are
$4.3\cdot 10^{-13}$~eV and $9.5\cdot 10^{-13}$~eV. It would be interesting to probe this  scenario with new data at 13 TeV.

\paragraph*{Acknowledgments}
We thank Dmitry Gorbunov and Konstantin Astapov for valuable discussions. We are grateful to Georg Weiglein for useful comments on the manuscript.
We are also grateful to Alexander Pukhov for his help with {\tt CalcHEP} code~\cite{Belyaev:2012qa} which allowed to cross-check results
obtained in this paper. The work was supported by the RSCF grant 14-22-00161. The numerical part of the work was performed on
Calculational Cluster of the Theory Division of INR RAS.

\end{document}